\documentstyle[12pt]{article}
\begin{document}
\bibliographystyle{unsrt}

\title{Kaon Productions Off Nucleons And The Structure Of Baryon Resonances}
\author{Zhenping Li\\
Physics Department, Peking University \\
Beijing, 100871, P.R. China}

\maketitle

\begin{abstract}
The recent investigations in the chiral quark model show that kaon productions
of nucleons play an important role in understanding the structure of baryons.
The evidences of a third $S_{11}$ resonance in the second resonance region
and two narrow states around 2 GeV suggest a set of the molecular type baryons
with the hidden strangeness. Confirming these states requires further
theoretical and experimental studies of the strangeness production. 
\end{abstract}

There have been considerable recent progress in the investigation of 
baryon resonances in the meson photoproductions. New experimental data\cite{eta}
for the $\eta$ photoproduction in the threshold region have been published.
These data, in particular the data from the Mainz accelerator MAMI, played
a very important role in studying the structure of the resonance 
$S_{11}(1535)$.  On the theoretical side, a new approach based on the
chiral quark model has been developed\cite{zpli97} for the meson 
photoproductions.  Comparing to other models at the hadronic levels, the
quark model approach to the meson photoproductions introduces the quark and
gluon degrees of freedom into the reaction mechanism, and it relates the 
photoproduction data directly to the spin flavor structure of baryon 
resonances.  Here, we would like to highlight some important features 
that one could learn from the forthcoming kaon production data at TJNAF
and the Bonn accelerator ELSA.

An important feature that we have learnt from the $\eta$ photoproduction 
off nucleons is the enhancement of the resonance $S_{11}(1535)$ and
 the suppression
of the resonance $S_{11}(1650)$ in the $\eta N$ channel.   
In Ref. \cite{zpli96}, we showed that this phenomenon can not be explained
in the framework of the constituent quark model. The solution of this problem
might come from the existence of a third $S_{11}$ resonance in the second 
resonance region. We indicated that there are 
considerable circumstantial evidences suggesting the presence of a
third $S_{11}$ resonance with mass 
$1.7\sim 1.8$ GeV and width around 0.2 GeV, which certainly can not be
accommodated by the quark model. Moreover, the data  from
SPHINX collaboration
in $p+C$ coherent diffractive productions\cite{sphinx} show 
two new states,
\begin{eqnarray}
X(2000)\to \Sigma^0 K^+, \quad M_{X(2000)}=1996\pm 7,\quad
\Gamma_{X(2000)}=99\pm 21 \nonumber \\
X(2050)\to \Sigma^*(1385) K^+, \quad M_{X(2050)}=2052\pm 6,\quad
\Gamma_{X(2050)}=35\pm 29 \nonumber ,
\end{eqnarray}
in which the masses and widths are in the unit of MeV.
 The resonances with such small
width at 2 GeV are unlikely to be normal $qqq$ states. Notice that
the threshold energies for $K^*\Lambda$ and $K^*\Sigma$ productions are
2007 and 2084 MeV respectively. A 
common feature for the $S_{11}$ resonances in the second resonance 
region and the states $X(2000)$ and $X(2050)$ is that
the masses of these states are just below the threshold energies of the 
kaon and $K^*$ productions.
This suggests that a set of molecular type baryons with hidden strangeness,
$K\Lambda$ or $K\Sigma$ state for the $S_{11}$ resonances in the second
resonance region, $K^*\Lambda$ for the $X(2000)$ and $K^*\Sigma$ for the
$X(2050)$,
may indeed exists bellow the threshold energies of Kaon and $K^*$ productions
in addition to the well known $\bar K N$ candidate $\Lambda(1405)$.
The $K\Lambda$ or $K\Sigma$ quasi bound state was first proposed for the
resonance $S_{11}(1535)$ in Ref. \cite{kl}, while the interpretation of 
the $K^*\Lambda$ and  $K^*\Sigma$ states for the $X(2000)$ and $X(2050)$
was suggested in Ref. \cite{shu}. In Ref. \cite{zpli96}
we pointed out that a pure $K\Lambda$ or $K\Sigma$ configuration for
the resonance $S_{11}(1535)$ is inconsistent with the data for the 
electromagnetic form factor of this resonance, and it should be strongly
mixed with the normal $qqq$ $S_{11}$ states.

How the $K\Lambda$ or $K\Sigma$ bound state is mixed with normal $qqq$ $S_{11}$
states has not been investigated theoretically. However,
the kaon productions would be very important channels to 
study the structure of these resonances experimentally.
Notice that the masses of the $S_{11}$ resonances are sandwiched 
between the threshold energies of the kaon and the $\eta$ productions,
the enhancement of the contributions from the $S_{11}$ resonances are expected
in the $\eta$ and kaon productions.
This is particularly true for the $\eta$ photoproduction in the threshold
region, where the dominance by the contributions from the resonance 
$S_{11}(1535)$ is well established.  A similar behavior should be expected
for the kaon photoproduction; our investigations\cite{mawx}
 in the $\gamma N\to K\Sigma$
reactions found that the resonance $S_{11}(1650)$ is indeed enhanced in
the threshold region of the reaction $\gamma n\to K^- p$.  
Thus, the presence of the third $S_{11}$ resonances could be tested
in the kaon production experiments. Moreover, as the new states $X(2000)$
and $X(2050)$ are just below the $K^*\Lambda$ and $K^*\Sigma$  threshold
energies, these states are expected to be enhanced in the reactions
$\gamma N\to K\pi \Lambda$ and $\gamma N\to K\pi\Sigma$, as there is 
no pomeron exchange to contaminate the cross sections. Unfortunately,
there are few data available for these reactions to either confirm or 
refute the presence of these two states.  Therefore, 
the experimental and theoretical studies of the $K\pi$ productions should
become one of the top priorities in the strangeness productions.
It is very interesting to note that the preliminary data from ELSA\cite{menze}
indeed have some hints of resonance structures in the 1.7 GeV region 
of the reaction
$\gamma p\to K^+\Lambda$ and around 2.0 GeV region in both reactions
$\gamma p\to
K^+\Lambda  $ and $\gamma p\to K^+\Sigma^0$, which might correspond to the
third $S_{11}$ resonance and the resonances $X(2000)$ and $X(2050)$. Of course
further experiments with better precision, in particular the data for the 
polarization, are needed.  Furthermore, the data in $\pi N\to K \Lambda$
or $\pi N\to K^+\Sigma$ are also required so that these new states could
be fully established.  

Thus, understanding the reaction mechanism of kaon productions has become 
increasingly important. There has been considerable recent progress in 
understanding the kaon photoproductions in the traditional isobaric 
models\cite{bijan} and in the newly developed chiral quark model\cite{zpli97}.
 An important  feature from  the quark model approach
 in the kaon production via $\gamma N\to K\Sigma$ is 
that the resonances $F_{37}(1950)$, $F_{35}(1905)$, $P_{33}(1920)$
and $P_{31}(1910)$ belonging to {\bf 56} multiplet in the quark model
play very important role in the reaction $\gamma N\to K\Sigma$.  
Better data in these reactions could provide us important 
information on the structure of these resonances as well.   

In conclusion, the high precision data for kaon photoproductions 
will provide us very important insights into the structure of baryon 
resonances that could not be possible from the pion photoproductions,
and establish the molecular baryons with the hidden strangeness.  
It may also help us to resolve the puzzle with the $S_{11}$ resonances
in the second resonance region. Without doubt, the strangeness
 productions will be an important and
exciting field in the near future. 

The author acknowledges the collaborations with R. Workman, Wei-Hsing Ma
and Zhang Lin on the works presented here. 
Discussions with  B. Saghai and R. Schumacher are gratefully acknowledged.
This work was supported in part by Peking University.

\end{document}